\documentclass[preprint,aps]{revtex4}
\usepackage[dvips]{graphicx}% Include figure files
\usepackage{dcolumn}% Align table columns on decimal point
\usepackage{bm}% bold math

\begin{document}

%\preprint{APS/123-QED}

\title{Indistinguishable Photons from a Single Molecule}

\author{A. Kiraz$^{1,2}$, M. Ehrl$^{1}$, Th. Hellerer$^{1}$\footnote{Current address: Experimental Physics, Chalmers University, 
Fysikgr\"{a}nd 3, SE-41296 G\"{o}teborg, Sweden}, \"{O}. E. M\"{u}stecapl{\i}o\u{g}lu$^{2}$, 
C. Br\"{a}uchle$^{1}$, and A. Zumbusch$^{1,3}$}

\affiliation{$^1$ Department Chemie und Biochemie and Center for Nanoscience, Ludwig-Maximilians Universit{\"a}t
M{\"u}nchen, Butenandtstr. 11, 81377 M\"{u}nchen, Germany\\
$^2$ Department of Physics, Ko\c{c} University, Rumelifeneri Yolu, 34450 Sariyer, Istanbul, Turkey\\
$^3$ Department of Physics and Astronomy, University College London, Gower Street, London WC1E 6BT, UK}

\date{\today}

\begin{abstract}
We report the results of coincidence counting experiments at the output of a Michelson interferometer 
using the zero-phonon-line emission of a single molecule at $1.4~K$. Under continuous 
wave excitation, we observe the absence of coincidence counts as an indication of two-photon interference. 
This corresponds to the observation of Hong-Ou-Mandel correlations and proves the suitability of the 
zero-phonon-line emission of single molecules for applications in linear optics quantum computation.

\end{abstract}

%\pacs{Valid PACS appear here}% PACS, the Physics and Astronomy
                             % Classification Scheme.
%\keywords{Suggested keywords}%Use showkeys class option if keyword
                              %display desired

\maketitle
Two otherwise indistinguishable photons arriving at a 50/50 beam splitter from different input channels, 
will interact and leave through the same output channel~\cite{grangier2002}. This phenomenon 
of two-photon interference is due to the bosonic nature of photons. Its observation is an important
demonstration of the quantum theory of light. Apart from its fundamental 
aspect, two-photon interference also lies at the heart of a broad range of applications in 
quantum information science. Photon-photon interactions mediated by two-photon interference allow for 
quantum gate operations employing photonic qubits. In such a scheme, linear optical elements, ideal 
photon counters, and true single photon sources were proven to be the only elements necessary to 
perform quantum computation~\cite{knill2001}. The requisite single photon sources for these applications should provide indistinguishable 
(i.e. exhibiting ideal two-photon interference) single photons on demand. In practice, the construction of
such single photon sources is one of the major tasks for the realization of linear optics quantum computation schemes. 
To date, experimental demonstrations relying on two-photon interference mostly used photons 
generated in a parametric down conversion process~\cite{hong1987}. However, with this photon source, the 
probability of simultaneous two-photon emission is governed by Poissonian statistics. Therefore, only low photon fluxes can be used. 
This limitation has been one important motivation for the recent development of a variety of true 
single photon sources based on different single two-level emitters 
~\cite{demartini1996,brunel1999,lounis2000,michler2000,santori2001,beveratos2002a,A.Kuhn:PRL02,J.McKeever:Sci04}. 
These single photon sources can emit single photons with a sub-Poissonian photon statistics provided that a large 
collection efficiency is realized. Only very recently, the coherence properties of the emitted single photons have been investigated.
Indistinguishability has been demonstrated for photons emitted by single quantum 
dots~\cite{santori2002,fattal2004a,fattal2004b} and single trapped atoms~\cite{legero2004}. 

Single molecules are an attractive alternative to both, single quantum dots and single trapped atoms. In self-assembled 
InAs quantum dots, the achievable coherence length is limited to $\sim750~ps$~\cite{borri2001,bayer2002}. By contrast, 
dephasing in single molecules is mainly due to phonon 
coupling which is suppressed at cryogenic temperatures. Nearly transform limited emission from single molecules 
with coherence lengths close to $4.9~ns$ is therefore readily observed using vibronic excitation at $1.4~K$~\cite{kiraz2004}. 
In comparison to single atoms, single molecules offer the advantage of much longer observation times. Experiments with 
one single molecule over more than 20 days have been reported~\cite{kulzer1997}. Single molecules have previously been employed 
as a source for triggered single photons using rapid adiabatic passage~\cite{brunel1999} and pulsed optical 
pumping~\cite{demartini1996,lounis2000}. Together with the large observed coherence times, these experiments 
prove the potential of single molecules as sources for indistinguishable single photons. In this letter, we 
report Hong-Ou-Mandel experiments using the zero-phonon-line (ZPL) emission from a single terrylenediimide 
(TDI) molecule. As a result of these experiments, two-photon interference is demonstrated.

Recently reported experiments on two-photon interference using single quantum 
dots~\cite{santori2002,fattal2004a,fattal2004b} and single trapped atoms~\cite{legero2004} all employed pulsed 
excitation schemes. In these cases, a Michelson interferometer with a path length difference that is exactly 
equal to the pulse separation is employed. This guarantees that consecutively emitted photons 
will simultaneously arrive at the two inputs of the beam splitter with a non-vanishing probability. 
Hence, depending on the coherence properties of the photons, two-photon interference is observed. 
In contrast, we have employed continuous wave (cw) excitation in our experiment. Under cw excitation, the path 
length difference between the two arms of the Michelson interferometer is not matched to an exact value. Rather, 
a path length difference which is larger than half the coherence length of the emitted photons is selected. This is 
sufficient to ensure that two independent photons with negligible magnitude squared overlap integrals~\cite{santori2002} 
will simultaneously arrive at two inputs of the beam splitter 
with a certain probability. The signature of two-photon interference is then revealed as the lack of coincidence 
counts at the beam splitter output.

The Michelson interferometer used in our experiments is depicted in Figure~1. 
We have used a rotatable $\lambda / 2$ plate in the longer arm of the interferometer in order to achieve 
parallel or orthogonal polarizations in channels 1 and 2. Considering parallel polarizations, a coincidence 
counting experiment between channels 3 and 4 reveals the normalized second-order coherence function:
\begin{eqnarray}
g^{(2)}_{34}(t,\tau) &=& \frac{\langle
\hat{a}_3^\dag(t)\hat{a}_4^\dag(t+\tau)\hat{a}_4(t+\tau)\hat{a}_3(t)\rangle}{\langle
\hat{a}_3^\dag(t)\hat{a}_3(t)\rangle\langle
\hat{a}_4^\dag(t+\tau)\hat{a}_4(t+\tau)\rangle} \,, \label{g2_nor}
\end{eqnarray}
$\tau$ represents the delay time between channels 3 and 4. 
In general, the expression of $g^{(2)}_{34}(t,\tau)$ in terms of $\hat{a}$, the photon annihilation operator of 
the molecular emission at the entrance of the interferometer, will be composed of 8 correlation functions and the 
corresponding complex conjugates. For this general case, the single photon wave functions in channels 1 and 2, 
described as $|\psi_{t_0}\rangle\propto\int^{\infty}_{t_0}dt e^{-\Gamma_{spon}(t-t_0)/2+i\phi(t)}a^{\dag}(t)|0\rangle$, 
are not orthogonal. $t_0$ denotes the initial time of the photon wave packet and $\phi(t)$ is a random function describing the pure 
dephasing process~\cite{santori2004,scully}. The magnitude square of the inner product of delayed and undelayed single photon wavefunctions, 
$\left|\langle\psi_{t_0}|\psi_{t_0-\Delta t}\rangle\right|^2$, can be shown to be proportional to $ e^{-2\gamma\Delta t}$ where 
$\gamma=1/T_2=\Gamma_{spon}/2+\gamma_{pure}$ is the total dephasing rate of the ZPL 
including spontaneous emission ($\Gamma_{spon}$) and dephasing due to other sources ($\gamma_{pure}$). 

For our case of interest, in the limit of large $\Delta t$ ($\Delta t \gg 1/(2\gamma$)) and 
small $\tau$ ($\tau \ll \Delta t$), Eq.~\ref{g2_nor} is largely simplified. In this limit, the two-photon wave 
function in channels 1 and 2 can be described as an outer product of two independent single photon wave functions 
$|\psi_{t_0}\rangle|\psi_{t_0-\Delta t}\rangle$ with $\left|\langle\psi_{t_0}|\psi_{t_0-\Delta t}\rangle\right|^2 = 0$. Thus, 
the delayed and undelayed operators will satisfy the commutation relationship:
$[\hat{a}(t-\Delta t+\tau_1), \hat{a}^\dag (t+\tau_2)]=0$, where $\tau_1$ and $\tau_2$ represent small 
delays (${\tau_1,\tau_2}<\gamma$) in the coincidence experiment. In the case of parallel polarizations at the exit of the 
interferometer, $g^{(2)}_{34}(t,\tau)$ is then expressed as~\cite{kiraz2004b,kiraz2004c}:
\begin{small}
\begin{eqnarray}
g^{(2)}_{34\parallel}(\tau) & = & \frac{1}{2} \left ( g^{(2)}(\tau) + 1 \right)
- \frac{\sin^2\theta\cos^2\theta}{\cos^4\theta + \sin^4\theta} \left|g^{(1)}(\tau)\right|^2 \label{g2_34}
\end{eqnarray}
\end{small}In this equation $g^{(1)}(\tau)=\frac{\langle
\hat{a}^\dag(t)\hat{a}(t+\tau)\rangle}{\langle
\hat{a}^\dag(t)\hat{a}(t)\rangle}$ and $g^{(2)}(\tau)=\frac{\langle
\hat{a}^\dag(t)\hat{a}^\dag(t+\tau)\hat{a}(t+\tau)\hat{a}(t)\rangle}{\left(\langle
\hat{a}^\dag(t)\hat{a}(t)\rangle \right)^2}$ correspond to the normalized first-order and second-order coherence 
functions of the ZPL emission of a single molecule respectively. Note that we drop the dependence of the coherence 
functions on $t$ due to cw excitation conditions. Transmission and reflection coefficients in the beam-splitter are noted 
as $\cos^2{\theta}$ and $\sin^2{\theta}$ respectively. For mutually orthogonal polarizations, photons at both input channels are 
completely distinguishable. For this case, the result 
of a coincidence experiment between channels 3 and 4, reveals: 
\begin{eqnarray}
g^{(2)}_{34\bot}(\tau) & = & \frac{1}{2} \left ( g^{(2)}(\tau) + 1 \right)\,.
\label{g2_34ort}
\end{eqnarray}

Vibronic excitation as used here relies on the excitation of a fast relaxing (relaxation time $\sim$1-10~ps) high energy vibrational 
level. In such a three-level incoherent excitation, the first and the second order coherence functions are given by analytical 
expressions as $g^{(1)}(\tau)  =  e^{-\gamma\tau}$ and $g^{(2)}(\tau)  =  1-e^{(W_P-\Gamma_{spon})\tau}$ \cite{kiraz2004b}. 
$W_P$ corresponds to an effective pumping rate. Using these equations, an examplary solution of $g^{(2)}_{34}(\tau)$ is 
depicted for parallel (solid) and orthogonal (dashed) polarizations. As Eqs. \ref{g2_34}, \ref{g2_34ort} assume 
small delay times, Fig. 2 is plotted for delay times between $-0.5/\Gamma_{spon}$ and $0.5/\Gamma_{spon}$. In the solid 
curve, the signature of two-photon interference is the absence of coincidence events around zero delay time. 

The vibronic excitation scheme is depicted in Fig. 3a~\cite{kiraz2003,kiraz2004}. In short, 
we used a narrow band cw laser excitation tunable around $605~nm$ (Coherent 899) to excite a vibronic transition of TDI 
embedded in the Shpolsk'ii matrix hexadecane. The samples were prepared by adding hexadecane to a solution of TDI in 
CHCl$_3$. CHCl$_3$ and O$_2$ were removed by several freeze and thaw cycles using liquid N$_2$ and intermediate 
evacuation steps. After this, the TDI/hexadecane solution was saturated with Ar. A drop of this solution was then 
quickly inserted into the precooled cryostat. Finally, the sample was cooled down to $1.4~K$ in a liquid He bath. An aspheric 
lens (NA=0.55) was used to focus the excitation and collect the emission light. The emission from the ZPL of a single 
TDI molecule was then filtered with a narrow band~(${\rm FWHM} = 1~nm$) interference filter before being focused onto a confocal 
pinhole (75~$\mu m$ diameter) placed inside a 4x telescope. In order to select single molecules from different spatial 
positions in the sample, a piezo scanner (Attocube ANP100) was used. Once a molecule was selected, its emission could 
be observed over several days. Fluorescence was either directed to a spectrograph (Jobin-Yvon HR460 with an Acton 
Research CCD camera, spectral resolution 30~GHz) or a homebuilt Michelson interferometer (Fig. 1). The path length 
difference between the two arms of the interferometer was equivalent to a temporal delay of $\Delta t = 4.6~ns$. One of 
its arms was equipped with a rotatable $\lambda /2$-plate. The output signals at the beam splitter were detected using 
avalanche photodiodes (APD, Perkin Elmer) in a Hanbury-Brown and Twiss configuration. The APD outputs were sent to a time 
to amplitude converter (TAC) as start and stop pulses. After adding an electronic delay on the stop pulse, the output of 
the TAC itself was stored in a multi channel analyzer (MCA). The resulting MCA histogram corresponds to the second-order 
correlation function $g^{(2)}_{34}(\tau)$ which is normalized using the function's value at large delay times. A 
temporal resolution of $420~ps$ was achieved.

A high resolution emission spectrum of the single TDI molecule is shown in Fig. 3b. A resolution limited 
band accompanied by the broad phonon sideband and vibronic emission lines is seen. The intensity ratio between 
the ZPL and all of its sidebands is governed by the Franck-Condon and Debye-Waller factors. In the emission spectrum in Fig. 3b, 40~\% of 
the single molecule emission intensity is orginating from the ZPL. This 
ratio can be further improved by placing the molecule inside a cavity~\cite{santori2002}. We investigated this 
molecule for five days including several cooling cycles between $1.4~K$ and $100~K$. All data presented in this work 
have been recorded from this molecule. Occasionally, spectral jumps of the molecule were observed during the 
experiment (Fig. 3c). Excitation of the molecule at the new spectral position always allowed to bring back the 
absorption to the initial position at 14988~$cm^{-1}$ within a few seconds. A similar switching behavior 
was reported previously~\cite{kulzer1997}. The digital spectral jumps are a clear signature of single molecule 
observation. In these rare events, the APD counts were observed to drop to $<5\%$ of the original level. Hence a signal 
to background ratio $>95\%$ was achieved in the experiments.

The result of coincidence measurements in which the ZPL emission of the single TDI molecule was sent into the interferometer 
and recombined at the beam splitter without any polarization changes is depicted in Fig. 4a. In this normalized 
correlation function, at the zero delay position, the value of $g^{(2)}(0)$ becomes 0.4. By contrast, Fig. 4b depicts the result 
of the coincidence detection experiment when rotating the polarization of the fluorescence in one interferometer 
arm by 90$^\circ$. Clearly, the resulting correlation curve exhibits higher $g^{(2)}(\tau)$ values around zero delay time. 
The normalized difference between the two curves is shown in Fig. 4c. In this curve, the peak around zero delay time 
constitutes the proof for the observation of Hong-Ou-Mandel correlations.
Figures 4d and 4e show the normalized difference between the data of Fig. 4a and another measurement with parallel 
polarizations, as well as for the data of Fig. 4b and another measurement with orthogonal 
polarization on the same molecule. In contrast to Fig. 4c, no signal above the noise level is seen at zero delay in both of these cases. 

Ideally, if the photons at the input of the beam splitter were completely indistinguishable, the value of $g^{(2)}(0)$ 
in Fig. 4a would be 0. Instead we observe $g^{(2)}(0)\sim 0.4$ which corresponds to a coincidence reduction factor, 
$V(0)=\frac{g^{(2)}_\bot(0)-g^{(2)}_{\parallel}(0)}{g^{(2)}_\bot(0)}$, of $V(0)=0.24$. 
The coincidence reduction factor can be lowered for several reasons. 
Imperfections in the mode matching of our interferometer are one source of 
contrast reduction. Because of its higher throughput, a conventional Michelson interferometer was preferred 
over a setup using a single mode glass fibre in our experiment. With a reduced interferometer path length 
difference of $1~ns$, we observed a fringe contrast of 0.7 for single molecule emission. However, 
due to the loss of single photon interference, we were not able to determine the fringe contrast for 
the $4.6~ns$ path length difference used in the correlation measurements, but the value of 0.7
is a upper boundary for this case. 
In addition to interferometer imperfections, spectral dynamics below the 30 GHz resolution of our monochromator 
could deteriorate the contrast. If these jumps 
occur on the ns time scale of the excited state lifetime, they will give rise to dephasing and
line broadening. While we did not determine the ZPL emission linewidth $\gamma$ of this molecule with a spectroscopic measurement, we
showed previously that stable 65 MHz emission is readily achieved in this sample system \cite{kiraz2004}.
Considering a typical spontaneous emission rate $\Gamma_{spon}\sim 1/3.4~ns^{-1}$\cite{kiraz2004}, a fit to the 
correlation measurements in Figs.~4a and 4b yields $\gamma_{pure}\sim 1/5~ns^{-1}$. Hence 
a photon coherence time of $T_2 \sim 3~ns$ is derived. For this 3~ns coherence time determined from our experiments 
and 4.6~ns path length difference, the magnitude squared overlap of the single photon wavefunctions 
in channels 1 and 2, $\left|\langle\psi_{t_0}|\psi_{t_0-\Delta t}\rangle\right|^2=e^{-2\gamma\Delta t}$, 
reveals 0.05 instead of 0. Thus, although our experiments clearly demonstrate the observation 
of Hong-Ou-Mandel correlations, Eqs. \ref{g2_34} and \ref{g2_34ort} are only valid to a good approximation 
due to the assumptions $\Delta t \gg 1/(2\gamma)$ and $\tau \ll \Delta t$. 
A complete solution of Eq. \ref{g2_nor} is necessary for a full theoretical explanation. It should be noted that 
a complete analysis will also explain the general appearance of the correlation functions depicted in Fig. 4 for 
large delay times ($\tau>1/\Gamma_{spon}$). Such an analysis should reveal the sidelobes apparent 
in Figs. 4a and 4b at $\tau=\pm \Delta t$.

In conclusion, we have demonstrated that the ZPL emission of single molecules can be used to produce indistinguishable 
photons. This complements earlier experiments using single quantum dots~\cite{santori2002,fattal2004a,fattal2004b} 
and single trapped atoms~\cite{legero2004}. Single molecules are an attractive alternative to these systems since 
they combine long coherence times with long observation times. We observed a coherence time of $\sim3~ns$ over the 
course of five days. The level of observed interference 
can be improved by coupling the single molecule emission into a glass fibre. This, however, necessitates an improvement 
of the collection efficiency. Experiments using specially designed low temperature optics with a high numerical aperture 
are currently pursued. Finally, the experiments presented here can easily be extended to include a pulsed excitation scheme. 
This is not necessary for the demonstration of two-photon interference but will be important for practical applications.

The authors thank K. M\"{u}llen for a gift of TDI and A. Imamo\=glu for helpful discussions. This work was supported by the 
Deutsche Forschungsgemeinschaft, SFB533, and the Alexander von Humboldt Foundation (A.K.). O.E.M. acknowledges support from 
T\"{U}BA/GEB\.{I}P award.

\bibliography{interference}
\bibliographystyle{apsrev}

\newpage
FIGURE 1, Kiraz et al.
\begin{figure}
\begin{center}
  \centerline{\includegraphics[width=8.5cm,angle=0]{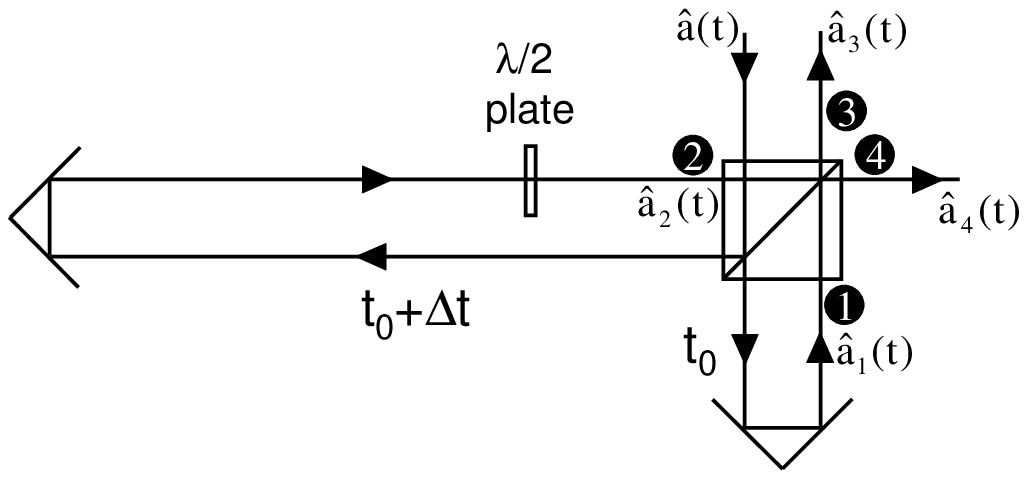}}
  \caption{Michelson interferometer used for two-photon interference measurements.}
\label{fig1}
\end{center}
\end{figure}

\newpage
FIGURE 2, Kiraz et al.
\begin{figure}
\begin{center}
  \centerline{\includegraphics[width=8.5cm,angle=0]{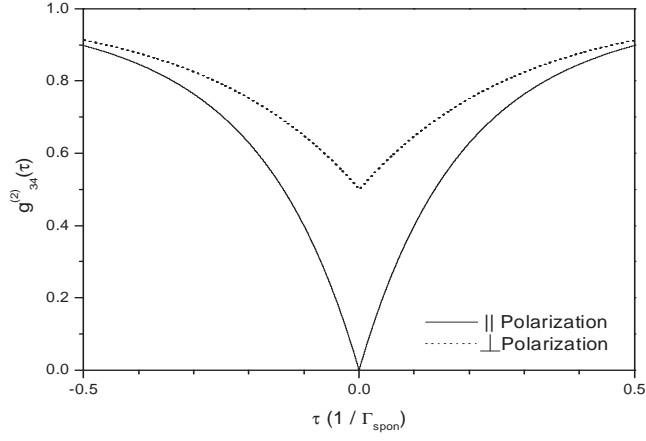}}
  \caption{Solution of $g^{(2)}_{34}(\tau)$ for large $\Delta t$ ($\Delta t \gg 1/\gamma$) and small $\tau$ considering 
  parallel and orthogonal polarizations. Parameter values are $W_P=2.5 \Gamma_{spon}$, $\gamma_{pure}=3 \Gamma_{spon}$, and $\theta=\pi/4$.}
\label{fig2}
\end{center}
\end{figure}

\newpage
FIGURE 3, Kiraz et al.
\begin{figure}
\begin{center}
  \centerline{\includegraphics[width=8.5cm,angle=0]{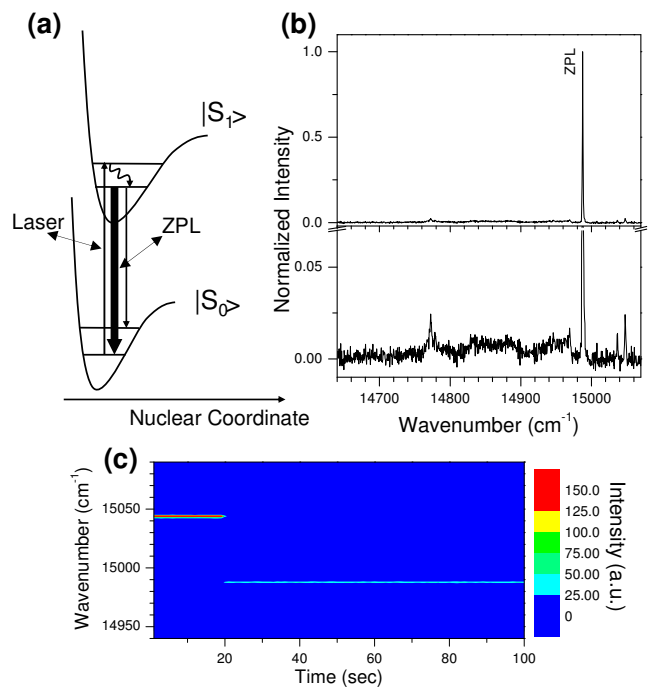}}
  \caption{(a) Vibronic excitation scheme. (b) High resolution emission spectrum of the TDI molecule. Two different 
  magnifications are plotted for clarity, excitation intensity is 4~mW.
  (c) Consecutive high resolution emission spectra under an excitation intensity of 1~mW.}
\label{fig3}
\end{center}
\end{figure}

\newpage
FIGURE 4, Kiraz et al.
\begin{figure}
\begin{center}
  \centerline{\includegraphics[width=8.5cm,angle=0]{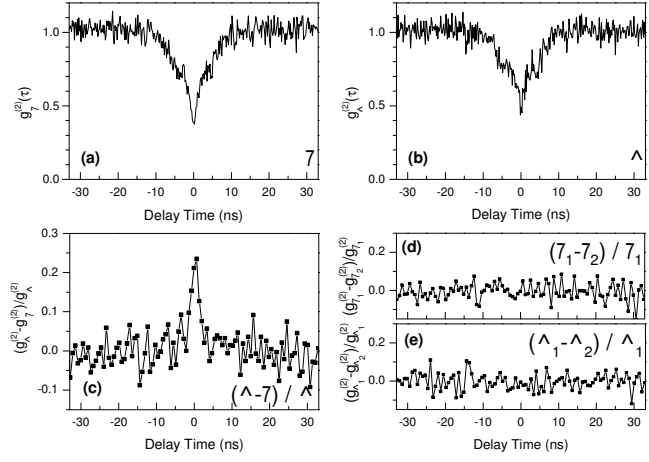}}
  \caption{Measured normalized second-order photon correlation functions for parallel (a) and orthogonal (b) 
  polarizations. (c) Difference between the correlation functions $g^{(2)}_{\bot}$ and $g^{(2)}_{\parallel}$ normalized 
  to $g^{(2)}_{\bot}$. Normalized difference between two measurements at parallel (d) and orthogonal (e) polarizations.
 Excitation intensity is 1~mW in a-e. Binning of the data is 2 in a and b, and 7 in c-e.}
\label{fig4}
\end{center}
\end{figure}

\end{document}